**Stability of Thin Film Refractory Plasmonic Materials Taken to High Temperatures in Air**

*Matthew P. Wells[1], Gomathi Gobalakrichenane[2], Ryan Bower[1], Bin Zou[1], Rebecca Kilmurray[1], Andrei P. Mihai[1], Neil McN. Alford[1], Rupert F. M. Oulton[3], Lesley F. Cohen[3], Stefan A. Maier[3], Anatoly V. Zayats[4] and Peter K. Petrov[1]\**

M. P. Wells, R. Bower, Dr B. Zou, R. Kilmurray, Dr A. P. Mihai, Prof. N. Alford, Dr P. K. Petrov
Imperial College London, Department of Materials, Prince Consort Road, London SW7 2BP, UK
E-mail: p.petrov@imperial.ac.uk
G. Gobalakrichenane
Polytech Paris UPMC, 4 Place Jussieu, 75005 Paris, France
Dr R. F. M. Oulton, Prof. L. F. Cohen, Prof. S. A. Maier
Imperial College London, Department of Physics, Prince Consort Road, London SW7 2AZ, UK
Prof. A. V. Zayats
King's College London, Department of Physics, Strand, London WC2R 2LS, UK

Materials such as W, TiN, and $SrRuO_3$ (SRO) have been suggested as promising alternatives to Au and Ag in plasmonic applications owing to their refractory properties. However, investigation of the reproducibility of the optical properties after thermal cycling at high operational temperatures is so far lacking. Here, thin films of W, Mo, Ti, TiN, TiON, Ag, Au, and SrRuO3 are investigated to assess their viability for robust refractory plasmonic applications. Films ranging in thickness from 50 - 180 nm are deposited on MgO and Si substrates by RF magnetron sputtering and, in the case of $SrRuO_3$, pulsed laser deposition, prior to characterisation by means of AFM, XRD, spectroscopic ellipsometry, and DC resistivity. Measurements are conducted before and after annealing in air at temperatures ranging from 300 - 1000° C for one hour, to establish the maximum cycling temperature and potential longevity at temperature for each material. It is found that $SrRuO_3$ retains metallic behaviour after annealing at 800° C, however, importantly, the optical properties of TiN and TiON are degraded as a result of oxidation. Nevertheless, both TiN and TiON may be better suited than Au or SRO for high temperature applications operating under vacuum conditions.

# 1. Introduction

Resulting from the resonant oscillations of free electrons within a metallic material upon stimulus by electromagnetic radiation, surface plasmons offer a powerful means of coupling the energy of light to the electrons at metal/dielectric interfaces, breaking the diffraction limit.[1][2] As a result, the field of plasmonics has been extensively researched in recent years with a view to the advancement of technologies such as sub-wavelength imaging,[1] communication systems,[3] photonics[4] and energy harvesting.[5]

Historically the noble metals Au and Ag have formed the basis of plasmonics research stemming from the pioneering work of Michael Faraday.[6] These materials offer several advantages, such as high electrical conductivity[7] and, in the case of Au, good chemical stability.[2] However, they are each subject to a number of practical limitations, such as significant interband transition losses,[8] incompatibility with standard CMOS technology,[9] and relatively poor thermal stability,[2] although the latter has been partially mitigated by a thin alumina coating.[10] The strong field confinement characteristic of plasmonic nanostructures inevitably gives rise to high electromagnetic energy densities, and hence the localised heating of devices.[11] The resulting lack of thermal stability is particularly detrimental for applications at high temperatures. For example, the area of solar thermophotovoltaics (STPV) stands to benefit significantly from the development of refractory plasmonic metals, offering the potential for energy conversion efficiencies of up to 85%.[12][13] However, in order to achieve such figures there is a need to find new plasmonic materials capable of delivering high absorption over a large frequency range, while withstanding the necessarily high operating temperatures, typically in excess of 730° C.[14][15] Additionally, refractory plasmonic materials are expected to facilitate advances in data storage technology, in particular the development of heat-assisted magnetic recording (HAMR) techniques.[16] Once again, such applications require the thermal stability of materials at high temperatures, as localised heating effects are an integral aspect of the HAMR technique.

While recent research has been conducted into the temperature dependence of the optical properties of Au and Ag thin films,[11] data regarding the long term thermal stability of thin films of the refractory materials W, Mo and Ti are scarce. It is also expected that the behaviour of materials experiencing high temperatures in air and in vacuum will be different. Although TiN[17][18] and, more recently, SRO[19] have been proposed as promising alternative plasmonic materials to Ag and Au owing in part to their refractory properties, experimental research in support of theoretical predictions is severely lacking. This article describes a comparative study between the temperature dependence of the optical and electrical properties of Au, Ag, TiN, TiON and SRO, along with known refractory metals Mo, W, and Ti. Experiments are conducted before and after annealing in air at high temperatures for one hour, where the temperature is raised incrementally to a maximum of 1000° C and slow cooled.

## 2. Results

Samples of W, Mo, Ti, TiN, Au and Ag were produced via RF reactive magnetron sputtering on (100) oriented MgO and Si substrates. Additionally, epitaxial SRO samples were produced using pulsed laser deposition (PLD), with MgO (100) as a substrate material. The thicknesses of TiN, TiON, Au, and Ag samples were found to be $50 \pm 5$ nm and SRO samples were measured as $95 \pm 5$ nm. Samples of W, Mo and Ti were 80, 130, and 180 nm respectively, with thicker samples having been produced to avoid surface oxidation dominating the measured optical and electrical properties. The XRD data for each material are provided and discussed in the Supplementary Information.

**Figure 1** shows examples of the AFM images before and after the anneal. The RMS roughness values for each sample are summarised in Table 1 (note that samples of Ti, Mo and W were measured over a $1 \times 1$ $\mu m^2$ area, while all other samples were measured over a $10 \times 10$ $\mu m^2$ area). By the 500° C anneal most films show significant topographic changes and enhanced roughness, while there is also a systematic increase in the roughness of the films after annealing at higher temperatures. For TiON and TiN, the increase in surface roughness is much less pronounced than the other films.

Figure 2 depicts the real and imaginary components of the dielectric permittivity for the noble metals, Au and Ag. Figures 3 and 4 meanwhile describe the dielectric permittivity for the pure metals, Ti, Mo and W, and for the metallic compounds, TiN, TiON and SRO respectively. Each of the three figures shows the results of measurements taken before and after annealing in air for 1 hour at temperatures between 300 and 1000° C. The optical data show a loss of metallic behaviour for each material. It should be noted that the pre-anneal room temperature optical measurements of Ag, Au, TiN, TiON and SRO[8][20][21][19] are found to be in agreement with previous studies. Additionally, the temperature dependence of the optical properties of Au appear consistent with the findings of Reddy et al.[11] However, in the case of TiN, the trends observed in the imaginary part of the dielectric permittivity are found to oppose those reported by Briggs et al,[22] while we find a transition to dielectric behaviour after annealing at 500° C which is in contrast to the high temperature stability (in excess of 1200° C) recorded by Briggs et al, where samples were heated under vacuum conditions. Meanwhile, both real and imaginary parts of the dielectric permittivity for samples of Ti, W, and Mo are found to be smaller in magnitude than those of reference data as a result of their propensity to oxidation, both during growth and upon removal from the vacuum chamber.[23][24] There is a correlation between the optical properties and the changes in the DC electrical resistivity, summarised in Table 2. Films annealed at high temperature show an increase in electrical resistivity.

Although the electrical behaviour of Ag, Au and SRO can be considered primarily a result of morphological changes, the electrical behaviour of TiN, TiON, and SRO, were considered in part a result of changes to the fundamental properties of the materials. Consequently, the electrical properties were examined in more detail, in particular, the changes in carrier concentration and mobility were extracted from Hall measurements, in combination with longitudinal resistivity, as a function of anneal history. These results are summarised in Table 3.

## 3. Discussion

The real and imaginary components of dielectric permittivity are described by the Drude equation,[25] where $\varepsilon_\infty$ refers to the background dielectric constant, $\omega_p$ to the plasma frequency, and $\Gamma_D$ to the Drude broadening.

$$\varepsilon'(\omega) + i\varepsilon''(\omega) = \varepsilon_\infty - \frac{\omega_p^2}{\omega^2 + i\Gamma_D\omega} \qquad (1)$$

For a perfectly free electron gas $\varepsilon_\infty$ may be assumed to be approximately equal to 1. Therefore, Equation 1 may be separated into real and imaginary parts described by Equations 2 and 3.

$$\varepsilon'(\omega) \approx 1 - \frac{\omega_p^2}{\Gamma_D^2 + \omega^2} \qquad (2)$$

$$\varepsilon''(\omega) \approx \frac{\Gamma_D\omega_p^2}{\omega^3} \qquad (3)$$

It should be noted that the above Drude model, although a reasonable model for Au and Ag in the upper visible and near-IR, does not account for the effects of interband transitions, and so is used here solely for descriptive purposes. The plasma frequency of a material is directly proportional to the charge carrier density, N, though inversely proportional to the effective mass, m*.

The Drude broadening is determined by electron – electron scattering, electron – phonon scattering, and surface scattering effects. An increase in surface roughness and, resultantly, in surface scattering effects may also increase $\Gamma_D$.

It can be seen from Table 1 and Figure 2 that Ag and Au behave similarly, with both real and imaginary parts of $\varepsilon$, along with surface roughness, increasing greatly above a given threshold temperature. It can be understood therefore that the increase in temperature is responsible for the measured increase in surface roughness,[11] and so too for an increase in the surface scattering effects represented by $\Gamma_D$. From Equation 2 and 3 it can be seen that an increase in Drude broadening will increase both $\varepsilon$' and $\varepsilon$''. Kim et al.[26] report that the resistivity of Ag becomes infinite above a given threshold temperature as a result of the agglomeration of the thin-film leading to a loss of film continuity. Furthermore, a thickness dependence of this threshold temperature is observed, with their results suggesting the cut-off temperature for a 50 nm Ag film to be in the region of 200° C. The results of Paulson et al. show that the degradation of Au can be understood in the same way.[27] The data presented here are consistent with these observations.

Although the refractory properties of Ti, W, and Mo are well reported in the case of bulk materials, Figure 3 shows the optical properties of thin-film samples are subject to significant variations at relatively low temperatures, with metallic behaviour being lost altogether after annealing in air at 500° C. In the case of Mo this can simply be attributed to oxidation, as confirmed by XRD measurements. By contrast, Ti and W are shown to undergo more significant morphological changes, as highlighted by Table 1. Furthermore, XRD measurements confirm a loss of crystallinity in the samples, leading to the observed loss of metallic behaviour.

TiN has been claimed as a good alternative material to the noble metals owing in part to its superior thermal stability.[17] Here however, changes are present in both $\varepsilon$' and $\varepsilon$'' after annealing in air at 300° C. It may be noted that the trends observed for $\varepsilon$' and $\varepsilon$'' in this case oppose each other, as $\varepsilon$' consistently increases with temperature, while $\varepsilon$'' decreases. As a result, it may be reasoned that the changes in optical properties arise from variations in $\omega_p$ rather than in $\Gamma_D$. This statement is supported by AC Hall effect measurements which show a decrease of a factor of 6 in the charge carrier concentration and an increase from 0.8 to 1.8 $cm^2$ $V.s^{-1}$ in charge carrier mobility upon annealing at 400° C, both of which may be expected to have the effect of raising $\omega_p$. After annealing at 500° C there is an increase in surface roughness which may be attributed to oxidation.

Such results appear contradictory to those presented by Li et al., who demonstrate that, in the context of a metamaterial absorber, the physical shape and optical performance of TiN remain unchanged after heating to 800° C for 8 hours, while the same conditions are sufficient to melt Au.[28] Moreover, Briggs et al show that TiN retains metallic behaviour in the visible regime up to temperatures in excess of 1200° C.[22] However, while the samples measured here were annealed in air, those of Li et al and Briggs et al. were annealed under vacuum conditions. Operation under vacuum limits the widespread applicability of TiN unless suitable encapsulation materials are identified; only under these conditions might one conclude that TiN is attractive for high temperature applications.

TiON can be seen to exhibit similar temperature -dependent properties to TiN. However, greater stability is observed between annealing temperatures of 300 and 400° C, and may be attributed to the fact the material is already partially oxidised. AC Hall effect measurements also show an improvement in stability, with charge carrier concentration changing only from $2 \times 10^{21}$ to $9 \times 10^{20}$ cm$^{-3}$ and mobility from 7.8 to 6.4 cm$^2$ V.s$^{-1}$.

The optical properties of SRO exhibit little change following annealing at temperatures below 800° C. However, some variations are observed at low frequencies, and both $\varepsilon'$ and $\varepsilon''$ show the same trend. In contrast to the behaviour of TiN, this indicates the dominance of $\Gamma_D$ in the temperature -dependent behaviour. This is supported by AC Hall effect measurements, which indicate that the charge carrier concentration remains approximately constant within the margins for error, decreasing from $5.5 \times 10^{22}$ to $4.3 \times 10^{22}$ cm$^{-3}$ while mobility remains at 0.1 cm$^2$ V.s$^{-1}$. After annealing at 1000° C however, a significant increase in $\varepsilon'$ is observed along with a decrease in $\varepsilon''$, suggesting $\omega_p$ to be the dominant variable; however, the large increase in surface roughness observed after this treatment may be expected to have the effect of increasing $\Gamma_D$. Indeed, it should be noted that the increase in surface roughness here is of a similar order to that of Au, as opposed to the less significant increase observed for TiN and TiON. Additionally, XRD measurements show the disappearance of peaks corresponding to (100) and (200) crystal orientations after annealing at 1000° C. It may therefore be concluded that structural and morphological changes following annealing at 1000° C result in the degradation of the optical properties of SRO, rather than oxidation, and so similar behaviour may be expected under vacuum conditions.

## 4. Conclusion

In summary, thin-film samples of eight materials, namely W, Mo, Ti, TiN, TiON, Ag, Au and SRO, have been produced through a combination of RF reactive magnetron sputtering and PLD in order to investigate their potential usage in high-temperature plasmonic applications. Measurements of the optical and electrical properties of each sample have found thin films of the commonly reported refractory materials W, Mo and Ti difficult to produce reliably, owing to the apparent influence of surface oxides. Meanwhile, the real part of the dielectric permittivity of Au was found to be stable up to 500° C when deposited on MgO, though the optical losses increased gradually with annealing temperature. After annealing at 600° C, significant changes to the morphology of the Au film were shown to result in a loss of connectivity across the film.

The optical properties of both TiN and TiON have been observed to change significantly upon annealing in air before a loss of metallic behaviour after annealing at 500° C. However, unlike the noble metals, changes to the material properties appear a result of oxidation rather than changes to the surface morphology. It may therefore be concluded that, while Au may be more suitable for high temperature applications in air, TiN and TiON are better suited to high temperature applications operating under vacuum conditions or after suitable protective capping layers are identified. Finally, the optical properties of SRO are considered, with the results showing that metallic behaviour is retained after annealing at 800° C. However, the degradation above this temperature appears a result of the agglomerate formation of the film.

Therefore, although SRO may be an attractive alternative to Au for some high temperature applications (if the latter lacks an alumina coating),[10] TiN and TiON may be expected to exhibit superior thermal stability in an encapsulated environment.

## 5. Experimental Section

All samples of Ag, Au, TiON and TiN were produced using a MANTIS deposition system via RF reactive magnetron sputtering. Titanium nitride samples were produced at 600° C using 200 W to achieve a deposition rate of 0.15 Å s$^{-1}$ in 30 % $N_2$ / $N_2$ + Ar atmosphere. Titanium oxynitride samples were produced at room temperature, again using 200 W to achieve a deposition rate of 0.15 Å s$^{-1}$ in 30 % $N_2$ / $N_2$ + Ar atmosphere. Ti, Mo and W samples were grown via DC magnetron sputtering at 200°C under an Ar atmosphere. For all samples, the sputtering chamber was pre-baked at 600°C for 2 hours followed by a Ti getter reaction to reduce the residual oxygen partial pressure to below $3 \times 10^{-9}$ mBar. A current of 300 mA was used for Ti and Mo samples and 280 mA for W samples to achieve a deposition rate of 1 Å s$^{-1}$. Samples of $SrRuO_3$ were prepared via pulsed laser deposition using a KrF excimer laser with a wavelength of 240 nm and pulse duration of 25 ns. The deposition process was conducted with a fluence of approximately 1.7 J cm$^{-2}$ and a repetition rate of 8 Hz under $O_2$ partial pressure of 50 mTorr. Once the deposition was complete the chamber was filled with $O_2$ to a pressure of 500 Torr and cooled to 30° C, at which point the sample was removed from the chamber. As with the samples produced by sputtering, the substrate material was double side polished MgO (100) with a surface area of 10 x 10 mm.

Samples were annealed in an MTI KSL-1100X-S furnace for 1 hour, the temperature having been raised over the course of 100 minutes. After annealing the samples were cooled to room temperature over a period of 1 hour.

All measurements subsequently conducted on the samples were conducted at room temperature ex-situ. Sample thickness measurements were conducted using a Dektak 150 surface profiler. DC resistivity was measured using the 4-probe technique. The optical constants of each sample were measured by means of ellipsometry. Measurements were conducted using a J. A. Woollam Co. HS-190 ellipsometer at an incident angle of 70°. To extract the data for each material the optical constants of the substrate were considered known, while the properties of the films were fitted directly to experimental data measured at 65 - 80° using the point-by-point method. X-ray diffraction measurements were acquired using a Bruker D2 PHASER system with a Cu Kα wavelength of 1.54 Å. A Lake Shore 8400 Hall system was used for AC Hall effect measurements with a field strength of 1.19 T and a current of 15 mA. AFM images were acquired using a Bruker Innova system operating in tapping mode with a scan rate of 1 Hz.

**Acknowledgements**

This work was supported by the Engineering and Physical Sciences Research Council (UK) Reactive Plasmonics Programme (EP/M013812/1) and by the Henry Royce Institute through EPSRC grant EP/R00661X/1. S.A.M. further acknowledges the Lee-Lucas Chair. P.K.P. and A.P.M. conceived and designed the research. M.P.W., G.G., R.K. R.B. and A.P.M carried out the experiments. All authors contributed to the manuscript writing and agreed on its final contents. The authors declare no competing financial interests.

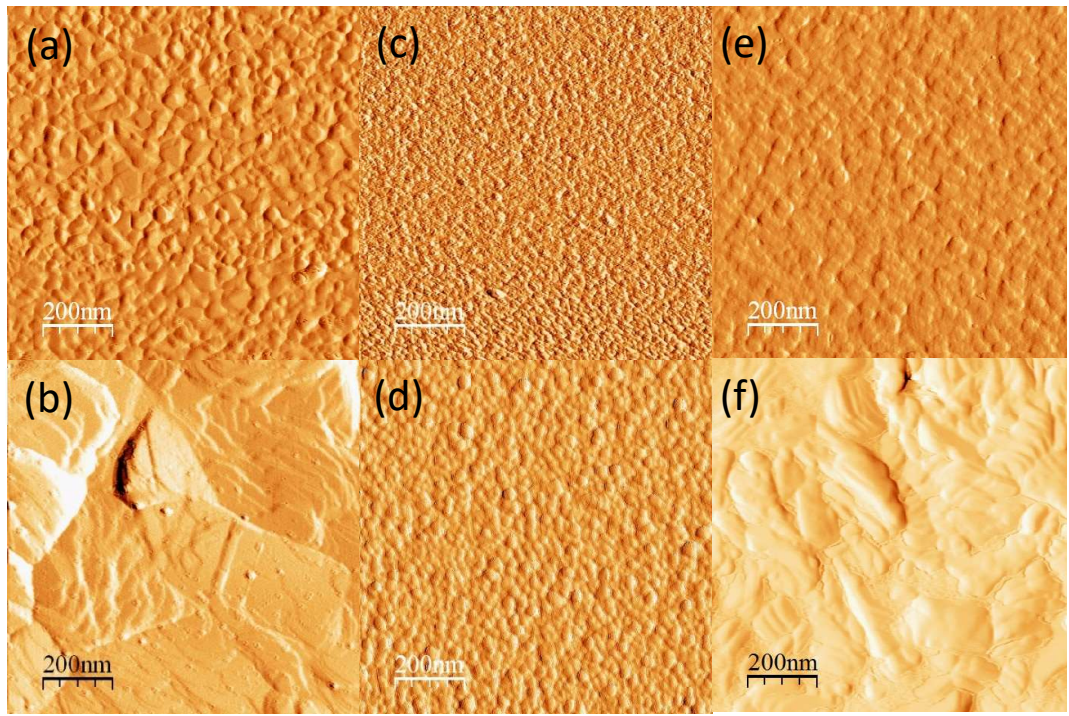

**Figure 1.** AFM images showing (a) Au before annealing, (b) Au after annealing at 600° C, (c) TiN before annealing, (d) TiN after annealing at 500° C, (e) W before annealing, (f) W after annealing at 500° C

| Annealing Temperature[°C] | RMS - Surface Roughness [nm] | | | | | | | |
|---|---|---|---|---|---|---|---|---|
| | Ti | Mo | W | Ag | Au | TiN | TiON | SRO |
| - | 9.2 | 2.8 | 1.1 | 2.4 | 1.2 | 0.7 | 0.7 | 1.4 |
| 300 | 9.8 | 3.8 | 1.5 | Not continuous | 0.9 | 0.7 | 0.7 | 4.4 |
| 400 | 2.3 | 4.2 | 0.8 | - | 0.7 | 0.5 | 0.5 | 7.5 |
| 500 | 3.5 | 2.1 | 10.3 | | 0.5 | 2.8 | 2.9 | - |
| 600 | | | | - | 1.9 | - | - | - |
| 800 | | | | - | - | - | - | 15.1 |
| 1000 | | | | - | - | - | - | 43.9 |

**Table 1.** Changes in surface roughness with annealing temperature

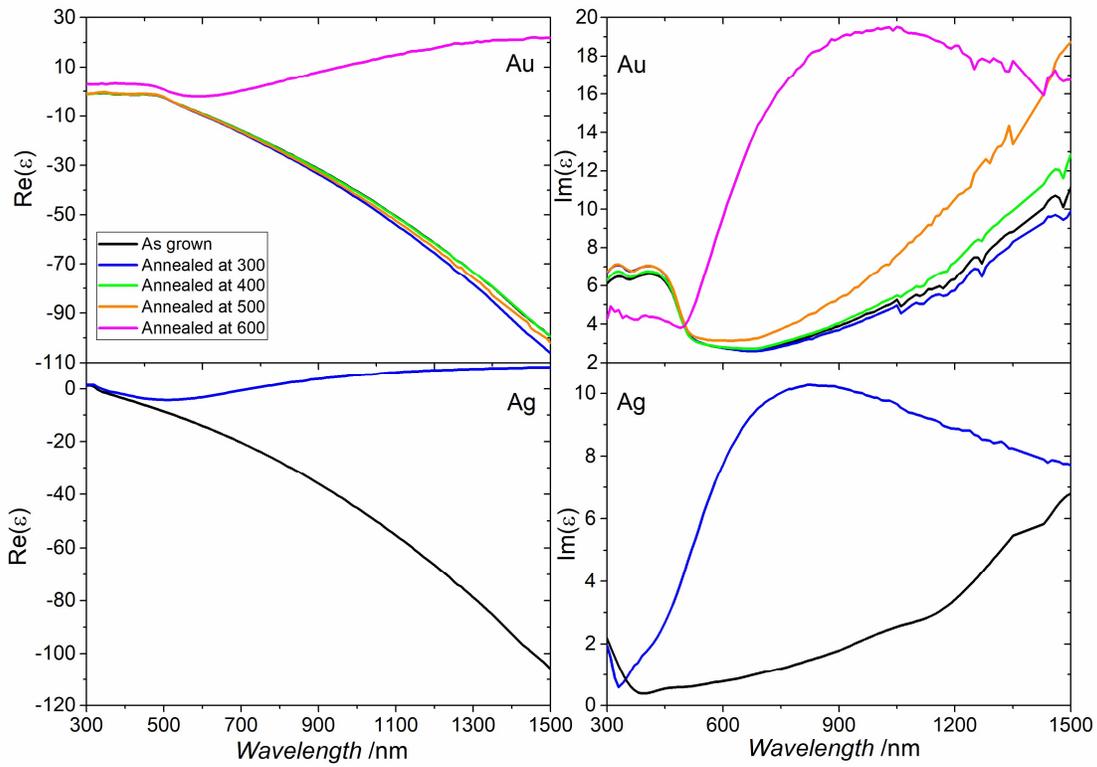

**Figure 2.** Ellipsometry data showing real and imaginary parts of the dielectric permittivity for samples of Ag and Au after annealing at temperatures from 300 - 1000° C

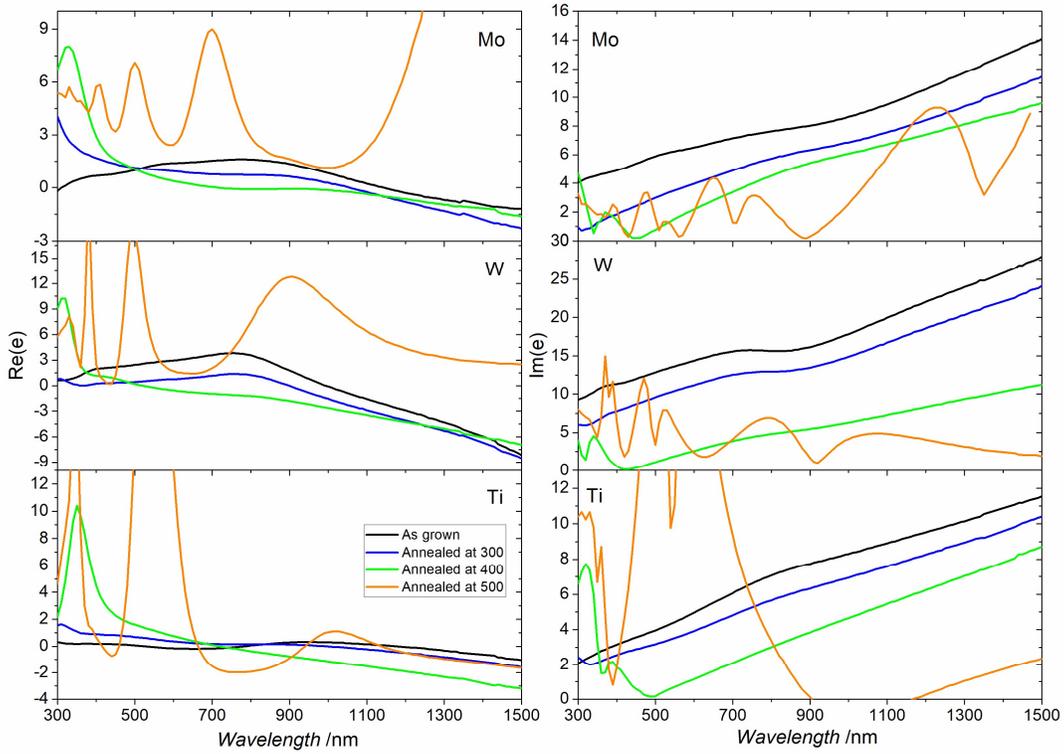

**Figure 3.** Ellipsometry data showing real and imaginary parts of the dielectric permittivity for samples of Mo, W, and Ti after annealing at temperatures from 300 - 1000° C

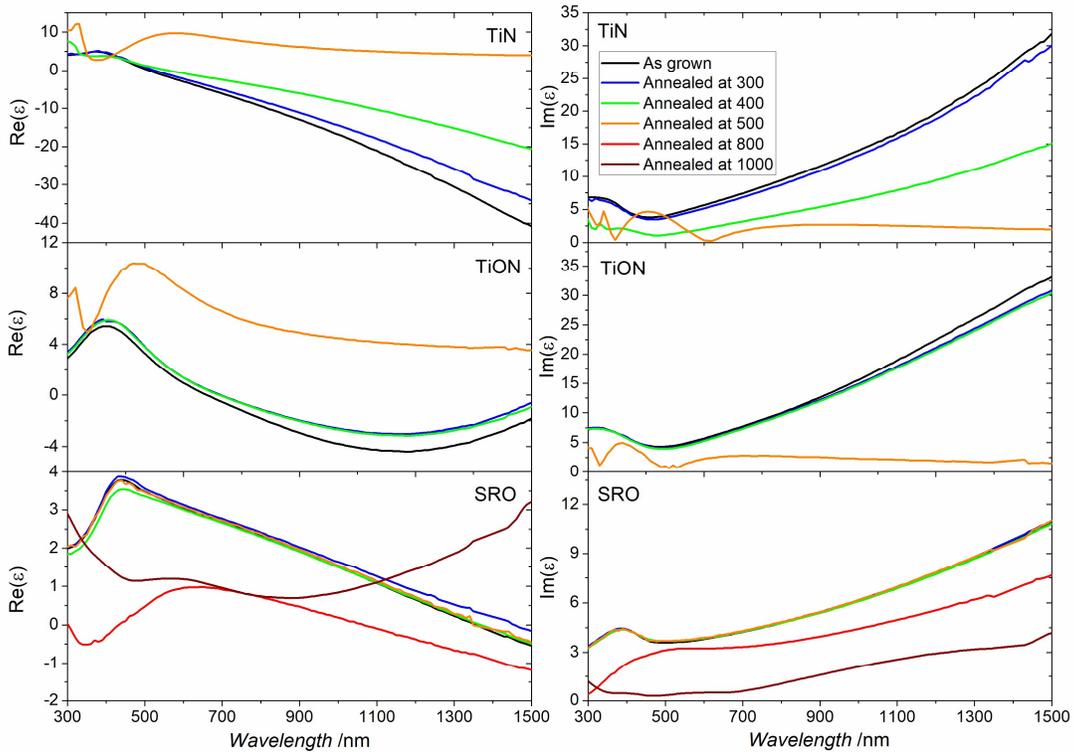

**Figure 4.** Ellipsometry data showing real and imaginary parts of the dielectric permittivity for samples of TiN, TiON and SRO after annealing at temperatures from 300 - 1000° C

**Table 2.** Changes in DC resistivity with annealing temperature

| Annealing Temperature [° C] | Resistivity [μΩ cm] | | | | | | | |
|---|---|---|---|---|---|---|---|---|
| | Ti [± 15%] | Mo [± 15%] | W [± 15%] | Ag [± 15%] | TiN [± 5%] | TiON [± 5%] | Au [± 20%] | SRO [± 10%] |
| - | 244 | 307 | 149 | 2.5 | 170 | 272 | 4.2 | 527 |
| 300 | 331 | 403 | 196 | ∞ | 177 | 295 | 3.6 | 590 |
| 400 | 534 | 380 | 182 | | 204 | 285 | 5.7 | 518 |
| 500 | 696 | ∞ | 5610 | | ∞ | ∞ | 3.7 | 514 |
| 600 | | | | | | | ∞ | - |
| 800 | | | | | | | | 569 |
| 1000 | | | | | | | | ∞ |

**Table 3.** Changes in charge carrier dynamics due to annealing

| | TiN | | TiON | | SRO | |
|---|---|---|---|---|---|---|
| | Room temperature | After annealing at 400° C | Room temperature | After annealing at 400° C | Room temperature | After annealing at 800° C |
| **Charge carrier density [cm$^{-3}$]** | $6x10^{22}$ $\pm\ 3\ x10^{22}$ | $1x10^{22}$ $\pm\ 5\ x10^{21}$ | $2x10^{21}$ $\pm\ 6\ x10^{20}$ | $8x10^{20}$ $\pm\ 4\ x10^{14}$ | $6x10^{22}$ $\pm\ 3\ x10^{22}$ | $4x10^{22}$ $\pm\ 2\ x10^{22}$ |
| **Charge carrier mobility [cm$^2$ V.s$^{-1}$]** | 0.8± 0.3 | 1.8 ± 0.7 | 7.8 ± 1.5 | 6.3 ± 1.3 | 0.1 ± 0.03 | 0.1± 0.03 |